\documentclass[MSNbibl,nameyear,dvips]{arxstspdf}
\usepackage{shere}
\usepackage{colortbl}
\usepackage{flushend}
\usepackage{stfloats}

\volume{27}
\issue{2}
\pubyear{2012}
\firstpage{308}
\lastpage{318}
\doi{10.1214/12-STS386} 

\makeatletter
\setattribute{abstract}{width}{36.5pc}
\setattribute{keyword}{width}{36.5pc}

\setattribute{abstract}{skip}{20\p@}
\setattribute{keyword}{skip}{8\p@}

\makeatother
\onecolumn
\begin{document}
\begin{frontmatter}

\title{After\hspace*{-0.4pt} 50$+$\hspace*{-0.4pt} Years\hspace*{-0.4pt} in\hspace*{-0.4pt} Statistics,\hspace*{-0.4pt} An\hspace*{-0.4pt} \mbox{Exchange}}
\pdftitle{After 50+ Years in Statistics, An
Exchange}
\runtitle{An Exchange}

\begin{aug}
\author[a]{\fnms{Jerome} \snm{Sacks}\corref{}\ead[label=e1]{sacks@niss.org}} 
\and
\author[b]{\fnms{Donald} \snm{Ylvisaker}\ead[label=e2]{ndy@stat.ucla.edu}}
\runauthor{J. Sacks and D. Ylvisaker}

\affiliation{NISS and UCLA}

\address[a]{Jerome Sacks is Director Emeritus of the National
Institute of Statistical Sciences,
PO Box 14006,
Research Triangle Park, North Carolina 27709, USA \printead{e1}.}
\address[b]{Donald Ylvisaker is Professor Emeritus, Department of
Statistics, UCLA
8125 Math Sciences Building, Box 951554,
Los Angeles, California 90095-1554, USA \printead{e2}.}

\end{aug}

%
\begin{abstract}
This is an exchange between Jerome Sacks and Donald Ylvisaker covering
their career paths along with
some related history and philosophy of Statistics.
\end{abstract}

%
\begin{keyword}
\kwd{Research areas}
\kwd{design of experiments}
\kwd{data}
\kwd{applied statistics}
\kwd{research programs}
\kwd{Statistics departments}.
\end{keyword}

\end{frontmatter}

{\fontsize{10pt}{\baselineskip}\selectfont
Jerome (Jerry) Sacks was born in 1931 in the Bronx. He graduated
from Cornell, with a 1952 B.A.
and a 1956 Ph.D. in Mathematics. His dissertation, with advisor Jack
Kiefer, was ``Asymptotic Distribution of
Stochastic Approximation Procedures.'' From 1956 until 1983 he taught at
CalTech, Columbia, Northwestern and Rutgers. In 1983--1984 he was Program Director
for Statistics and Probability at NSF. He returned to Academia as
Head of the Department of Statistics at the University of Illinois,
until 1991, when he became Professor at Duke. At the same time Sacks
became the founding Director of the National Institute of Statistical
Sciences, a position he held until 2000. When he stepped down the NISS
Board of Trustees established the Jerome Sacks Award for
Cross-Disciplinary Research to honor Sacks' service. In 2004 he retired
from Duke. Sacks is a Fellow of the IMS, the ASA and the AAAS,
and a recipient of the Founders Award of the ASA. During and after
his work at NISS Sacks studied highly complex scientific problems
such as circuit optimization, traffic simulation and air pollution
measurement, using both design strategies and computer models.

\hspace*{10pt} Donald (Don) Ylvisaker was born in 1933 in Minneapolis. His B.A. in
Mathematics and Economics was from Concordia College in 1954, followed
by an M.A. in Mathematics from the Unversity of Nebraska in 1956 and a
Ph.D. in Statistics from Stanford in 1960. His dissertation, with
advisor Emanuel Parzen, was ``On Time Series Analysis and Reproducing
Kernel Hilbert Spaces.'' From 1959 until 1968 he taught at Columbia, New
York University and the University of Washington. He then moved to
UCLA where he was head of the Division of Statistics in the
Department of Mathematics until his retirement in 1996. Ylvisaker has
served both the IMS and the ASA on many committees and in many
functions. He has done major editorial work for several of the leading
statistics journals. Between 1990 and 2008 he was involved with
advising the Commerce Department on Census adjustment and evaluation,
he is a~long-term advisor of state lotteries, and he has been involved
in projects counting the homeless population and in the sensible use of
DNA evidence in the criminal justice system. At UCLA he was
instrumental in the 1998 establishment of a Department of Statistics,
separate from Mathematics. Ylvisaker is a Fellow of the IMS and the
ASA.\looseness=-1

\hspace*{10pt} The conversation reported below is not a unique event. Sacks and
Ylvisaker have been friends and collaborators for a long time, with a
very distinguished list of joint publications, written over more than
30 years. Perhaps the most influential ones have been the papers on
design aspects of regression problems, which started in classical
mathematical statistics and eventually came to include calibration,
response surfaces and computer experiments. As documented in the
conversation below, we see the emphasis in the publications of both
Sacks and Ylvisaker shifting from more theoretical topics, such as
stochastic approximation and reproducing kernels, to papers using a
more applied and computational approach, which are motivated directly
by actual advice and consultation.

\begin{flushright}\vspace*{10pt}
 Jan de Leeuw

 September 2011
 \end{flushright}}

\twocolumn
\textbf{JS}: We met in 1959 at the Department of Mathematical Statistics at Columbia.
You had just arrived from Stanford as a fresh Ph.D.; I had been there for
two years.
I don't think Departments of Mathematical Statistics exist anymore in
the U.S.
There are a couple in Australia and England, and maybe in some places
nobody hears about.
The Columbia department morphed into a Department of Statistics and, by
now, there is a whole
swarm of names in use: Statistical Science, Statistical Sciences,
Mathematics and Statistics,
Statistics and Operations Research and who knows what else. I'm not
sure how the name
Mathematical Statistics came about, but the signage certainly suggested that
``data enter at their peril; theory is done here---for applications go
elsewhere.''

\textbf{DY}: We were trained as mathematical statisticians certainly (though I
had managed two
summers at Allison Division of General Motors, involved with data on
the X-ray determination
of stress in metals). In fairness to those times, there was a lot of
interest among the mathematically
oriented in fresh areas with considerable practical importance:
reliability, queueing theory, inventory
problems, flood/insur\-ance~risks and the like. Still, data didn't have
much of a~presence; the
thinking was closer to ``suppose there is a person with these data and
this problem, we will solve this problem.''
It was really a question of matching for a time---getting theorists
together with practitioners who had significant data issues.

So, maybe serious treatment of data was not pro\-minent in our circles until
we were middle-aged, but this is not to say that all such issues are
absent today.
Recall that when we were looking at nominations for the Mitchell Prize under
the standard of ``an outstanding paper where a Bayesian analysis has
been used to
solve an important applied problem,'' one could safely discard quite a
few methodological
works that had little direct connection with an honest problem.
Too commonly one finds papers that propose a new technique and then
tout its
performance on a data set rescued from another time or place. Still,
everyone has to
operate at some remove from the data, lest there be nothing with which
to go public.

\begin{table}
\noindent
\begin{tabular}{@{}>{\columncolor[gray]{0.92}}p{230pt}@{}}
{\fontsize{11pt}{12.8pt}\selectfont{\begin{tabular}{@{}p{230pt}@{}}\vspace*{3pt}
\centerline{\uppercase{\textbf{\textsf{Columbia in 1959--1960}}}}\\
\it
 \quad The regular faculty were Ted Anderson (chair), Howard Levene, Herb Robbins
and Lajos Takacs. Anderson had been there since 1946, as had Levene; Robbins
arrived in 1953 from the North Carolina, following a year at the
Institute for Advanced Studies.
In 1959 Takacs came from England, a way station following the Hungarian
revolution.
Other faculty appointments were Ron Pyke, Jerry Sacks and Don
Ylvisaker, while
Joe Gani and Harold Ruben were, ostensibly, visitors. More widely at Columbia,
Rosedith Sitgreaves was at Teachers College and Cy Derman was in
Industrial Engineering.

 \quad The department had offices scattered over three floors of Fayerweather
Hall, abutting
Amsterdam Avenue at 117th Street. Helen Bellows handled the entire
administrative
load and did the technical typing as well. Full-time students may have
had a
common room, but they mostly appeared for classes and seminars. Among
them were
Ester Samuel-Cahn, Gideon Schwarz, Joe Gastwirth, Ted Matthes and
Lakshmi Venkateraman.
There was, as well, a healthy traffic in ``night school'' students,
notably Peter Welch, who came down from IBM.

 \quad The younger faculty and visitors interacted a~great deal, in and out of
seminars, often joined
by Benoit Mandelbrot, Y. S. Chow and Dave Hansen from IBM.
Short-term visitors (Kai-Lai Chung and Aryeh Dvoretzky) and seminar
speakers (Alan Birnbaum,
Tom Ferguson, John Hartigan, Cuthbert Daniel) added to an already
spirited atmosphere.
Most memorable was Sir Ronald Fisher: as cantankerous as rumored and
into the tobacco/cancer debate.

 \quad The level of activity centered around the department that year led to
several long-term alliances,
that of Chow and Robbins, for instance. At the end of the year,
following a variety of misunderstandings
between senior and other faculty concerning personnel and future plans,
Pyke went to the
University of Washington, Sacks headed to Cornell and then
Northwestern, Ylvisaker went to
NYU and then Washington, Gani returned to Australia, and Ruben became
head of the statistics
department at the University of Sheffield. The wholesale exodus seemed
an unfortunate outcome at the time.
\end{tabular}}}\vspace*{3pt}
\end{tabular}
\end{table}

\textbf{JS}: Of course applications and data were the stuff of concern for many
in those years---sampling
was always there, serious quality control problems were being attended
to, designs for engineering
and agricultural experiments were on the table, as well as many
other
issues in, and especially outside,
academic circles.\vadjust{\goodbreak} But in the rarified climate of Cornell Mathematics
(where I got my degree),
Stanford Statistics (where you got yours) or Columbia Mathematical
Statistics (where we met),
what mattered more was the ability to advance the basic theory of
statistical reasoning.

I think that reflected the post-World War II era, a time when
structure, ambiguity and abstraction
were major forces guiding intellectual movements. Even if the dynamics
of abstract expressionism,
jazz, ``chance-composed'' music of John Cage, beat poetry, theater of
the absurd
or new wave cinema might not fit neatly in that box, their
characteristics of randomness,
uncertainty and subjectivity might resemble those forming the
attraction and development of statistics.
Statistics was caught up in efforts to seek structure (Wald's \textit{Theory of Statistical Decision
Functions} in 1950, Savage's \textit{Foundations of Statistics} in 1954),
while, of necessity,
pursuing the ``jazz'' of data analysis. It would be nice if some
intellectual historian
could explore and analyze these connections---I'm not equipped to do that.

The 1950s were exciting times for those of us who came to ``life''
then. The tension
between theorists and practitioners, new departments and expansions
driven in part by
Sputnik and the (overly optimistic) hope that decision theory would
resolve all philosophical
(and practical) disputes helped foster an environment that enabled
statistics to flourish.
Jack Kiefer's optimal design paper in the 1959 \textit{JRSS}, along with
the ensuing discussion and rejoinder,
provides an interesting snapshot of that statistics world.

\textbf{DY}: For me, it was coming to ``life'' in the micro\-cosm of the times
that was statistics at
Stanford~in the mid-1950s. Statistics was then regarded with some
interest by mathematicians
for its game theory and probability connections (Sam Karlin came~to
statistics for a while, for example),
as well as~by~econ\-omists and others (Kenneth Arrow and Pat Suppes were
often seen around the Stanford department, for instance).
These were heady, energetic times for Statistics, suggestive of an era
of great progress.~Yet these good feelings seemed to flag in the early 1960s; overall
respect for statistical problems
waned as\break \mbox{mathematical} statistics was found too hard and items like
Inventory Theory were rather easily ``resolved.''

\textbf{JS}: Math departments seemed eager to hire statisticians in the 1950s,
albeit the more theoretically inclined.
Certainly the increasing demand for teaching statistics was a factor,
and the proximity of interests in statistics
and probability at that time was another. Though this alliance of
interests weakened in subsequent years,\vadjust{\goodbreak}
it provided a measure of acceptance for statisticians within
mathematics then (after all, Kolmogorov was
everybody's ``daddy''), and a number of prominent
probabilist/mathematicians dabbled, and more, in statistics,
for example, Joe Doob, Mark Kac, Kai-Lai Chung and Sam Karlin.

\textbf{DY}: Whatever nuances one places on the research interests of the era,
and despite the excitement
generated by its seminal results, it is now a time that seems not all
that well remembered for its people.
Erich Lehmann died recently at 92, and there are long-established
researchers around who have not
much sense of him and his work, as just one example. Perhaps it was the
timing of serious, innovative
statistical work in the post-war years that brought out what were to us
the huge personalities of the 1950s;
one can compile a pretty long list, and one had the feeling that there
were many chiefs and not so many
Indians around. While their personae remain vivid to those around at
the time, statistics has now gone
off in so many directions that there are now few ``giants'' to be
readily discerned.

(As a footnote to research in the 1950s, I heard a~computer scientist
give a talk the other day in which
sufficiency and Rao-Blackwell entered without further ado---and we
thought those topics were
goners after data analysis and robustness came to the fore.)

\begin{table}
\noindent
\begin{tabular}{@{}>{\columncolor[gray]{0.92}}p{235pt}@{}}
{\fontsize{11pt}{12.8pt}\selectfont{\begin{tabular}{@{}p{235pt}@{}}\vspace*{3pt}
\centerline{\uppercase{\textbf{\textsf{The 1960 Berkeley Symposium}}}}\\
\it
 \quad
 For six weeks in the summer of 1960,~an~ex\-traordinary group of
statisticians and probabilists
met at the fourth of six symposia, held at five-year intervals at UC-Berkeley.
This symposium marks a high point of the widespread interest in the
more mathematical
aspects of statistics and probability. In the preface Jerzy Neyman
notes that
``the present Proceedings are much richer than those of the earlier
Symposia because of~the
several contributions from members of the great Russian school of probability.''
In four volumes, the Proceedings of the Symposium contained~over 100
works, including
such classics as~``Nonincrease, Everywhere of the Brownian Motion
Process,'' by Dvoretsky, Erdos and Kakutani,
and ``Estimation with Quadratic Loss,'' by James and Stein. Linking
via\texttt{
\href{http://www.lib.berkeley.edu/math/services/symposium.html}{http://www.lib.berkeley.edu/math/}
\href{http://www.lib.berkeley.edu/math/services/symposium.html}{services/symposium.html}}
details the dimension and character of all the symposia and speaks to
the special nature of the fourth one.

 \quad Of countless unrecorded memorable moments at the meeting, the comment
by Harold Hotelling following Doug Chapman's lecture on ``Statistical
Problems in Dynamics of Exploited Fisheries Populations'' stands out.
Hotelling gave a~lengthy, erudite exposition, ``Fish, as Symbol'' with
stress on the mythic
and religious to complement the secular content of Chapman's talk.
\end{tabular}}}\vspace*{3pt}
\end{tabular}
\end{table}

\textbf{JS}: Looking back to those times makes me reflect on how (I'm afraid to
ask why)
we got interested in statistics and what influenced our directions. In
my case I was
an undergraduate mathematics major and became curious about statistics
from an
offhand remark of my brother who had come into contact with Cuthbert
Daniel while
working at Oak Ridge and was impressed enough to suggest that
statistics was onto
something (little did I know). Then, in my senior year, Kiefer and Jack
Wolfowitz joined
the faculty at Cornell and, between course work and paper grading, I
became more involved
and was encouraged to stay on as a graduate student.

The mathematics department at the time was not very large (maybe 20--25 faculty).
There were two statisticians (Kiefer and Wolfowitz), a few probabilists
(Chung, Gil Hunt and Kac)
and fewer students (Bob Blumenthal and I were the only first-year
students interested in statistics or probability;
Dan Ray was finishing his Ph.D. at the time). The small and close
atmosphere in the
department had two effects on me: it forced a fair amount of
independence on me, and it
provided a strong intellectual influence. At the same time, I shouldn't
slight the fact that there\vadjust{\goodbreak}
was a strong group of statisticians across the Cornell campus, some of
whom had been there
before I started graduate work (Walt Federer, Iz Blumen and Phil
McCarthy), others who had
just arrived or were visiting (Bob Bechhofer, Charlie Dunnett, Lionel
Weiss and Milt Sobel)
and probably others that I fail to recall. It was a pretty intoxicating
atmosphere with such a
variety of statisticians around and an array of year-long or summer visitors
(Feller, Kakutani, Bochner, Dvoretzky and Erd\"os), but mainly was so
to me because of
the dynamism of Wolfowitz and Kiefer.

\textbf{DY}: In my case, it was always natural to take mathematics courses when
in school;
everything else seemed mundane by comparison. While shoring up my undergraduate
background in a master's program at the University of Nebraska, I
gravitated to Fred Andrews,
a recent statistics Ph.D. from Berkeley who had also spent some time at
Stanford. All to the good,
he got me to work hard and then encouraged me\vadjust{\goodbreak} to go on to a Ph.D. program.
In those days (1956), one applied to North Carolina, Berkeley and
Stanford, and then had a
choice among them. I was taken by the thought of heading west to
Stanford and became part of
their first large statistics class---ten full-time students came to
campus that year (Bill Pruitt and an older
Frank Proschan among them), joining two continuing full-time students---Don Guthrie and Rupert Miller.

Student camaraderie, an engaged and approachable faculty, streams of
visitors and related faculty
passing through Sequoia Hall, with its trafficked corridors and
unpretentious offices, contributed to the
exciting place Stanford was in those years; you can well imagine the
effect this had on a~student
from a~small Minnesota college. The lively research topics were
sequential design (Herman Chernoff),
admissibility (Charles Stein), total positivity (Karlin), and
reproducing kernel spaces and time series
(Manny Parzen). Manny agreed to take me on as his first student, and I
made it through school.

The serious mathematics/theoretical statistics\break training we got served
the two of us well throughout our
careers. I have always attributed the ``Mathematics as a secret
weapon''
thought to Art Owen---a~point well
made. At the end of the day, the whole Stanford experience was great
for me, but then it was time to set
out for life as a ``grown-up,'' to Columbia in the Fall of 1959.

\textbf{JS}: It is interesting that it was the combination of our backgrounds
that led in 1959 to our
collaboration: you were close to the innovations by Manny Parzen in
formulating time
series analysis, and I was aware of the seminal work of Kiefer and
Wolfowitz on optimal experimental designs.
It sure didn't hurt to have had the mathematics training that enabled
easy communication between us,
especially of the function space ideas.

\textbf{DY}: I recall that you gave a seminar on Kiefer and Wolfowitz's Annals
regression design paper,
from the June '59 issue, at the start of the school year, and I~wondered aloud to you about the
possibility of doing something related when errors were correlated.
Guess we got past that question after some 15 years.

\textbf{JS}: One of the first reactions to your question was in thinking about
how we might
optimally sample a Brownian motion. I don't think we concluded much at
the time, but,
after a few feeble starts, we managed to come to grips with the issues
in the early 1960s.\vadjust{\goodbreak}
By then both of us had left Columbia---you were in Seattle (U. of
Washington's Mathematics
Department) and I was in Chicago (Northwestern's Mathematics
Department). I continue to be
surprised that we were able to make any headway on the problem at all,
and even more
surprised that the asymptotically optimal designs we found for
polynomial regression with
Brownian motion errors were intimately connected with optimal designs
for numerical integration.
This even gave us some street cred in the applied math/numerical
analysis world.

We carried on this collaboration at long distance and with some visits
(mostly you to Chicago).
Sometimes the distances were extreme, with me working on the beaches in
Acapulco while
you were ``sweating'' it in Seattle. Later on I was ``amused'' when the
well-known mathematician
Stephen Smale had research funds withheld by the NSF for saying he did
his best work on
the beaches of Rio (actually, I think it was his outspoken support of
the Free Speech Movement
at Berkeley and the attention of the House Un-American Activities
Committee that led to the loss of funds).
Still later, in 1985, Smale had an encounter with some numerical
analysis problems (dubbed Computational Complexity
by Joe\break Traub and company) and rediscovered some of our results on
numerical integration
in an article in the \textit{Bulletin of the American Mathematical Society}.
I~sent him a note along with one of our reprints; he never answered.

\textbf{DY}: It was indeed surprising to make progress with the (infill)
asymptotics of our
design problem. You taught me a lot in that process and, it can be
added, forced
me to work awfully hard when it came to many parameter extensions.
The structural things that I knew more about, such as the connection of
splines with
Brownian Motion (or, more generally, processes with their kernel
sections), were hardly deep,
but it took a long time to understand that they could be posed in a way
that would be interesting to people.
Thus, a short distance from our regression problems to quadrature, but
several years before
we wrote it down in that fashion, and more years before quadrature
surfaced as one of the basic problems
in what had come to be called complexity theory.

Getting back to the state of Statistics in the late 50s, expansion
showed up in various ways.
IMS meetings were in those days, for example, written up in the News
and Notices section of the
\textit{Annals}, with the full list of attendees. Imagine attempting that
for a JSM today. There is an interesting
history in how\vadjust{\goodbreak} departments emerged and grew, one that Alan Agresti and
Xiao-Li Meng are now compiling at \texttt{\href{http://www.stat.ufl.edu/\textasciitilde aa/history/}{http://}
\href{http://www.stat.ufl.edu/\textasciitilde aa/history/}{www.stat.ufl.edu/\textasciitilde aa/history/}}.

By 1960 the mathematical statistics of the 1950s had lost some of its
attractiveness, but
Tukey's call to data analysis, and the related follow-up in the form of
the more theoretical
robustness questions sparked a new path for the 1960s. None of this was
entirely new, but who
could forget Tukey's talk of 1961 and paper of 1962, or Huber's thesis
of 1964? Are there other landmarks
of the decade if one sticks to the central thread we are on? True,
there were deep admissibility results
that continued the ``statistics as math'' thing, but I have the sense
that people were
searching for things to do with themselves after the basic theoretical
problems that remained
were found to be too tough.

\textbf{JS}: Looking back at the 1960s, I get a sense that while the world
around us was blowing apart
(civil rights movement, Vietnam war, assassinations of\break John and Robert
Kennedy and Martin Luther King),
the profession and its activities were moving independently in the way
you described with, I think, one exception:
Bayesian thinking was emerging more strongly and creating some tension
with frequentists.
I didn't feel much of this, possibly because I had less sympathy with
``fundamentalism'' and more interest in
responding to questions raised in the context of existing or developed
structures, such as those we addressed
in the work on designs when errors are correlated. At any rate, I had
trouble in determining who was in
charge of fixing the prior distribution.

\textbf{DY}: I have always had trouble with being lectured on the right way to
think, and the
Bayesian evangelists of the 1960s were very active. I have no problem
with Bayesian ascendancy
(do you recall the review we got some years back, to the effect that
``It's nice to find an intelligent Bayesian paper again\,\ldots but\,\ldots''), yet am stuck in the belief that
the statistical issues in a problem precede a
philosophical/methodological stance on its treatment.
Some situations, generally highly complex settings, demand priors (Toby
Mitchell persuaded me of that in his gentle,
nonpedantic way), but my applied work has most often been close to
sampling and design,
and correspondingly far from posing\break a~need for, or justification of, a
prior distribution.

Thinking ahead then to the 1970s and beyond, life in research and
teaching broadened for
each of us; there were added administrative jobs and more involvement
with applied statistics.
While this began to get serious in the early 1980s,\vadjust{\goodbreak} the question I~would raise is, did something
happen in the 1970s? Early on, at least, these were ``between'' years
for me, personally and professionally.
Perhaps it was also something of a forgettable decade for statistics generally.
By its end, one could point to Empirical Bayes, the wholesale onset of
smoothing problems in
their modern guise, the Bootstrap and, more generally, to the
increasing rumblings of the
computing revolution; a quiet interlude nonetheless?

\textbf{JS}: I suppose there are many who will argue that a~lot happened, but I
agree with you that the
decade of the 1970s was more important as a prelude to the more
explosive developments
of the 1980s. It is convenient to claim so as a generalization of what
was happening
personally, but I think, rather, that it was a more general phenomenon.

In the mid-1970s I was drawn into statistical issues directly spurred
by applied problems.
I suspect the estrangement between reality and my work was due to
living in a mathematics
department, or maybe was a hangover of the spirit of the 1950s. In any
case, it was around
then that I came into contact with Bob Boruch and Don Campbell, social
psychologists at Northwestern.
Interestingly, it was Cliff Spiegelman, getting his Ph.D. in the math
department, who pulled me into that
milieu and into Campbell's work on quasi-experimental and regression
discontinuity designs.
(Never underestimate the power of an imaginative graduate student to
light the paths of senior professors.)
Though I never published anything directly, Spiegelman did. You and I
also produced some work on
smoothing methods that grew out of these problems. The more significant
thing, for me at least, was
being part of a conversation that stimulated my thinking (and maybe
some of theirs) about the critical
statistical issues faced in evaluating social policies and innovations
(e.g., Head Start).
It also opened me up to influences from people who had modest technical
expertise,
yet had an incisive intuitive understanding of the statistical nature
of their data.

\textbf{DY}: With age comes wisdom, or the times demanded it? I got involved
with legal
work, lottery consulting, census and various other matters that became
increasingly
interesting to me and worth the time spent.

\textbf{JS}: Practice seemed to me to come in (at least) two forms. There were
applications
within the scientific research world, and others that stemmed from
sources like those
you mentioned. The former applications were relatively easy to
transition to, in principle,
but the others brought different issues that depended very much
on
personality and politics.
I~did get caught up in some employment discrimination cases, and later
had an extensive
involvement with voting rights cases in the 1980s. Sorting out and
explaining statistical subtleties,
or even crudities, to a mixed bag of intelligent but quantitatively
semi-literate clients, lawyers and judges,
while being challenged and scrutinized by opposing experts of varying
degrees of sophistication,
forces one to have a firm and critical view of just what statistics is
about. It's comparatively easy
to prove a theorem by imposing the right assumptions; it's another
story to justify assumptions
to a suspicious antagonist or decision maker. I suspect we can regale
each other with stories of ``experts''
unable to do elementary arithmetic, judges willing to admit
probabilities of 1.14, lawyers engaged in
Bayesian dialogue and so on. (The last named actually took place in a
deposition in a voting rights case:
Sam Issacharoff, an able lawyer and currently a professor at the NYU
Law School, fenced with me
about why I wouldn't do or accept a Bayesian analysis. I forget who won.)

\textbf{DY}: I always thought that a principal reason for being involved with
legal matters was the
need to keep Bayesian analyses out of the courts. Mannered subjectivity
was even to be
foisted on jurors in the form of ``choose your own prior probability''
so that the proffered
crank could be turned---evangelism of the 1960s now brought forward
for the masses.

In the 1980s, statistical testimony was regularly offered as to
questions of
discrimination by race, gender and age in such areas as employment, wages,
housing, jury selection, sentencing and voting rights. I was involved
in several cases
during this time and, yes, once in a voting rights case with David
Freedman, Steve Klein and you.
In that instance, the opposition employed an ecological regression that
had Stockton's
diverse citizenry composed of two politically cohesive racial groups---blacks and Hispanics on the
one hand, whites and Asians on the other!

The atmosphere changed somewhat with the landmark 1987 decision in
\textit{McCleskey v. Kemp} in
the wake of the Baldus death penalty study, for it brought heightened
standards for ``statistical'' relief:
the ``ra\-cially disproportionate impact'' in Georgia death\break penalty
sentencing, indicated by
a comprehensive scientific study, was not enough to overturn the guilty
verdict without
showing a ``racially discriminatory purpose.'' One commentator had it
as ``the Dred Scott decision of our time.''
However viewed, the statisticians' discrimination landscape underwent a
considerable change.\vadjust{\goodbreak}

Another active area began in the 1970s with the reporting of blood and
tissue typing
tests as evidence of culpability. DNA analyses were then a leap forward
in this same vein when introduced
in 1987, shortly after they became available. Statistics enters the
discussion only through population genetics,
and in a rather cursory fashion. The tests, as evidence in court,
brought out a fierce battle when DNA
analyses were first invoked, prematurely in my opinion. This was played
out quite publicly in the 1995
O.~J.~Simpson trial. Things subsided a good deal after the second of two
NRC reports on its use, in 1996.
Unfortunately, to me, common sense had lost out and the ``product
rule'' is now practiced with little added
thought; it likely will continue to be used until genetic advances
finally eliminate the need for silly calculations.

\textbf{JS}: The changes in attitude and involvement with practical issues we
both experienced left me,
by 1980, restless and dissatisfied with the same old same old academic
pursuits in a mathematics department.\break
I~began looking around to change circumstances, and things moved at a
rapid pace.
Pivotal for me was a~decision to go to the NSF as a program director in 1983.

Jack Kiefer's untimely death in 1981 had had a~profound effect. Not
only was he a close friend
but a~man who shared his ideas and encouraged others to take on new challenges.
He had previously suggested that I go to the NSF to take a~hand in
advancing the cause of Statistics in
Washington. After his death I thought more about doing so, and in the
early spring of 1983, Ingram Olkin and
Don Rubin pressed me to take on that job. One thing that drove me was
the perceived chance to
affect the future development (read funding) of statistical design of
experiments. With Kiefer's death the
leading figure in the field was gone and it was unclear how and in what
way future efforts would proceed.
In fact, as you may recall, at the Neyman--Kiefer Symposium in 1983, and
again at the annual JSM
meetings in the summer of 1983 in Toronto, several of Jack's friends
and collaborators
(Ching-Shui Cheng, Toby Mitchell, Henry Wynn, you and~I) talked about
where the field was going.
None of us saw a strong direction at that time and we thought it
valuable to pursue ways to
energize thinking about this. You and Ching-Shui followed up by putting
together a proposal
to stage a series of four workshops on design.
These were funded by NSF and held at Berkeley and UCLA in 1984--1986.

\textbf{DY}: It is hard to properly account for Jack's influence on us all, but
one could
start by bringing out his ideas and technical strengths, his personal
warmth and generosity.
He managed the combined role of mentor and friend with remarkable
grace, and there we were,
lost for both his leadership and his companionship.

Working with Ching-Shui, laconic but with much to say, on the planning
and implementation of the
workshops, was a new experience for me, and a joy. The first, held at
Berkeley in the summer of 1984,
brought together researchers with a fairly broad spectrum of interests;
the summer workshop in 1985
at UCLA hosted a truly wide array of interesting people (among them
Rosemary Bailey, Grace Wahba
and Don Rubin) and topics (climate research, survey design and
nonresponse, and product and
process design for manufacturing, as examples).

\textbf{JS}: The last workshop in January 1986 was an important one. In fact, at
the end of the
workshop I enlisted Henry Wynn, over sushi at a restaurant near the
UCLA campus, to help
draft a research proposal to formulate and attack problems on
statistical issues in computer
experiments. This helped start a whole program of research at Illinois
and elsewhere---the workshop
had some real influence.

\textbf{DY}: It was the most focused of the four workshops. In setting up the program,
I was able to rely on Toby Mitchell, who had been thinking of these
things for some time.
This was my first opportunity to work with (and appreciate) Toby, and
the resulting program
was an early and distinguished entry in what was soon to become a
central research area.

\textbf{JS}: There is something, less obvious, to be learned from that experience.
The NSF was, probably still is, most often regarded as a source of
funding for ideas
generated within the discipline. What is perhaps less noted is the
catalytic effect of NSF;
the stimulus provided by NSF to you and Ching-Shui is a nice example of that.

\textbf{DY}: It does seem that the NSF is presently far more involved in the
pushing of broader
research agendas than in the years prior to your tenure there---thus,
cross-disciplinary
areas might be identified for specific grant monies as opposed to the
classic meth\-od of
soliciting individual research proposals. Surely, as with upstream
design for a manufacturing process,
this is a sensible method for shaping and facilitating research
programs. In this vein, it is crucial
that statistics is suitably championed and, in the complex and shifting\vadjust{\goodbreak}
statistical research environment
one currently sees, that its sub-areas come under a wise focus. Easier
said than done.

\textbf{JS}: The year at NSF put me on another trajectory: I~went to Illinois in
1984 to lead the
establishment of a new department of statistics, and also became
increasingly involved in
subject matter issues. A lot of my time at NSF was spent with
scientists from outside statistics
and mathematics and I began to sense, along with some others, that our
field could and should
be energized by serious interest across disciplines. This was not fully
appreciated by all, but it did
resonate at UIUC with some enthusiasm for joint appointments and
enterprises. Also, in 1984--1985,
I helped start a study about cross-disciplinary research in statistics
that led ultimately to a
recommendation of an Institute devoted to that purpose.

\textbf{DY}: Coincidentally, 1984 marks an effort toward cross-disciplinary
statistics at UCLA that
began with a proposal by social scientists to hire some six
statisticians in their division.
The statisticians then located in the mathematics department sought
some involvement in
that process, at the very least. The dean of Physical Sciences was not
sympathetic to us,
but the dean of Social Sciences shortly set up a statistics division in
the social sciences program,
one that brought some cohesiveness to the process. Most importantly,
Jan de Leeuw was recruited
to run it from a joint appointment in Psychology and Mathematics. Much
energy having been generated,
a division of statistics was formed within the Mathematics Department
in 1986, leading eventually to the
birth of the Statistics Department in 1998. In this last development
the (then) dean of Physical Sciences
was highly instrumental, being persuaded that Statistics was an honest
endeavor of great interest to
many in the university. Of course, he was right.

\textbf{JS}: The cross-disciplinary theme emerged more gradually as an influence
in the
1980s when compared with that of computational developments. Both
continue to underpin
attitude and focus in the field. Strangely though, some advances in
computational power,
like supercomputing, were slow to be recognized by our colleagues, and
the rapid, innovative
adoption of statistical methods and ideas by computer scientists (and
others) was not quickly
digested. To a degree, this gave rise to some thinking of the need to
push the field.
I became especially aware of these things when I was at the NSF.

\textbf{DY}: I was no monitor of the changing times, certainly, but can offer
before and after
pictures from UCLA. When I was a vice chairman of undergraduate affairs
in the math
department in 1971--1973, we proposed an undergraduate degree in applied
mathematics to
sit alongside the one ``pure'' math option available to students.
The proposal was promptly laughed out of the faculty meeting, probably
without a vote being taken,
for it would have allowed some students to graduate without a
differential geometry course!
There were only two or three applied faculty to defend or implement
things at the time,
and they had to vie with the slightly more numerous statistics group
for respectability in
the department. It was several years before such an applied major was
instituted.

Fast-forward to the present to find just over 200 majors in each of
pure and applied math,
and 90 students enrolled in the new undergraduate statistics major.
Since the 1970s, the applied mathematics group in the Math Department
has grown
considerably, is awash with money and prestige, and is now ranked about
third in the country.
The Statistics Department dates to 1998, the FTE count has roughly
doubled since then,
and the student population has gone through the roof.

Of particular note, there has been a considerable movement of
mathematicians into
problems we have thought of as statistical, at least to some degree.
For example,
for the years 2009--2011 at the Institute for Pure and Applied Mathematics
at UCLA,
one finds programs on ``Model and Data Hierarchies for Simulating and
Understanding Climate,''
``Mathematical Problems, Models and Methods in Biomedical Imaging,''
``Statistical and Learning-Theoretic Challenges in Data Privacy,'' and
``Navigating Chemical Compound Space for Materials and Bio Design.''
Some statisticians have shown up in these programs, but not many.

Of course the big news is, and should be, the data themselves: huge
increases in availability,
much improved recognition of the need for the understanding of basic
statistical concepts in ``everydata''
problems, and the astonishing growth of analytic tools mindful of new
age data sets and
rapid computational improvements.

In this expansion, the design and analysis of computer experiments has
been a
special interest of ours. The early framework papers that grew out of
the workshops in the
1980's, already with an eye toward various engineering problems, are
now heavily
cited in many areas in which the simulation of complex systems is practiced.
One would like to think that the ideas in them, and beyond citations,
are put to use in the kind
of experiments that get written up as internal company or laboratory
reports on specific projects.
Since the late 1980's you've been a~lot closer to the ``factory floor''
in this regard than I~have.

\textbf{JS}: You are right to point out that the development of computer experiments
coincided with the attention to cross-disciplinary work.
Statisticians\break
don't ``own'' computer
models and dealing with computer experiments means collaborating with
the subject
matter people who use the models. It was natural to be engaged
simultaneously with the
computer experiment research and the efforts that led to the
establishment of NISS
with its mission of fostering and doing cross-disciplinary statistical research.

\textbf{JS}:
When I look back at the history of NISS's creation, I am struck by the
number of
leaders of our field, and outsiders as well, willing to engage in and
support such a venture.
Of course there wasn't unanimity, but the story does reflect a
willingness of leadership to
push boundaries despite low odds of eventual success (even in
retrospect, investing with
Bernie Madoff might have been a safer bet). That characteristic, surely
not unique to our field,
may have some roots in our having to claw our way into the
consciousness of established
authorities (see your experiences at UCLA and everybody's everywhere else).

\textbf{DY}: I suppose only the older persons among us have the time to fret
over the status and
stature of statistics, the young are hard at work on ``doing it.''
Still, the pushing of institutions
like NISS needs incisive goals, thorough planning and plenty of clout.
There, it seems, one needs age, experience and foresight.

\textbf{JS}: And a measure of luck. Little happens from just plain intention---help is needed from
many\break
sources. We typically focus on the advances in the intellectual arc of
statistics and pay
less attention to the politics affecting us and others. The ``local''
politics exemplified in the
creation of NISS is minor compared to the connections statistics has
with the serious
economic, social and political matters of our time. These connections
need much more attention.
There are some books and occasional articles, but I don't think they
capture the bigger
and critical picture of what we are about.

\begin{table}[t]
\noindent
\begin{tabular}{@{}>{\columncolor[gray]{0.92}}p{230pt}@{}}
{\fontsize{11pt}{12.8pt}\selectfont{\begin{tabular}{@{}p{230pt}@{}}\vspace*{3pt}
\centerline{\uppercase{\textbf{\textsf{Origins of NISS}}}}\\
\it
 \quad At the 1984 annual IMS meeting in Tahoe, California, discussions
about the
future of the field among David Moore, Ingram Olkin, Ron Pyke, Jerry Sacks,
Bruce Trumbo and Ed Wegman led to a plan for a report on cross-disciplinary
research in statistics. Money was obtained from the NSF, and a panel
was formed
with Olkin and Sacks as co-chairs. At a meeting in 1987 Olkin proposed the
establishment of an Institute to implement ideas around which the panel
had coalesced;
in time, the proposal became the key action item in the report.

 \quad How to bring the recommendation to reality began with discussions among
Nancy Flournoy, Olkin, Sacks and, most critically, Al Bowker. These
discussions led,
with the help of Flournoy and Murray Aborn (NSF), to the financing of a
feasibility
study carried out through the ASA, culminating in a plan to seek
proposals from
groups around the country (mostly located in the East).
Proposals competed not for dollars, but to receive blessing from a committee
of statisticians (chaired by Bowker) the proposers had to commit real
dollars themselves!

 \quad A consortium from the Research Triangle area of North Carolina made the winning
proposal, committing start-up money, academic positions, land and funds
for a building.
A host of North Carolina people were involved: university provosts,
department chairs,
executives at the Research Triangle Institute and others. Two people
were critical for
the initial effort and for the early stages of growth of NISS. One, Dan
Horvitz, had
stature in the statistical world and, as retired vice-president at RTI,
had significant political contacts.
The other, Sherwood Smith, CEO of Carolina Power \& Light, had great
interest in furthering
the development of Research Triangle Park, and his political savvy and
connections were instrumental
in ensuring the initial commitments for NISS and, a few years later, a
renewed commitment by the
state to build a ``house'' for the institute.
\end{tabular}}}\vspace*{3pt}
\end{tabular}\vspace*{1pt}
\end{table}

\textbf{DY}: A political case in point is the census adjustment controversy that
began in the
early 1980s, peaking over the 1990 and 2000 censuses. The
pro-adjustment view was\vadjust{\goodbreak}
the dominant one: a viable method was in place to better the process of
counting every person.
There was a good deal of informed opinion in this direction, and much
uninformed support in
the statistical community---the capture/recap\-ture story is readily
recounted, but its use in the
census context is far more complex than that. The other side
emphasized, among other things,
the heterogeneity of the post-strata that were central to the
adjustment methodology. The discussion of the
technical issues was greatly complicated by the Demo\-crats' support of
adjustment and the
Republicans' opposition to it.
The media had a field day over the matter, and little of benefit
accrued to the Bureau of the Census,
or to Statistics.

My own involvement with the Census Bureau---contracts, contacts and
NRC panels---lasted close to twenty years, beginning with the 1990 census. I was in
the camp that held the
nonadjustment decisions of the 1990 and the 2000 Censuses to be proper;
in the latter case,
the bureau agreed with that position at the last moment, especially
given problems with duplicates.
In all, the more one is around the bureau, the more respect one has for
the tasks it is given,
and for the host of talented people who work toward its goals.

\textbf{JS}: The census issues, along with the DNA and voting rights experiences
we talked
about earlier, hit a nerve. I don't think it too idealistic to want
statistics to appear in these
contentious settings as objectively as possible. The rush to employ
sophisticated,
or not so sophisticated, methods under tacit assumptions---Hardy-Weinberg in DNA calculations,
ecological regression in voting rights, independence of
capture/recapture in census adjustment---that may lack adequate justification is harmful, even when used to
advance laudable causes.

\textbf{DY}: We do have a PR problem at all levels. The much-improved
early training of students in probability and statistics
notwithstanding, reaching the public is
not easily managed. There is the constant barrage of social science
findings, medical recommendations
and the like. Of late, the often-fleeting nature of study results gets
a good deal of attention,
but it is hard to see how the system and the media will ever reward
patience in such matters;
we show up as would-be custodians of the peace in these settings,
sharing the fate of
commentators on rare illnesses, earthquakes, climate change and the
like.\looseness=1

\textbf{JS}: Books like \textit{How to Lie With Statistics}, \textit{Fooled~by
Randomness} and
\textit{The Black Swan}
(at first I thought the last was a late review of an old Tyrone Power,
Maureen O'Hara pirate movie---a movie with much more pertinence to the
economic catastrophes of 2008
than Weibull distributions)
too often leave\break a~sense of villainous activity by statistical
practitioners.

Just what can be done to further a nonwarped public perception has been evasive.
It is impossible to shut off the supermarket tabloids. And while useful
efforts to bring some
public sense to vexing reports like those about mammogram screening
have appeared in
such places as the \textit{New York Times}, none of the journals or
newspapers of record has
actually undertaken to spotlight the ubiquitous nature of uncertainty
and the efforts to cope with it.
Individual instances pop up now and again but a~coherent discussion,
perhaps in a series of articles,
would be useful.

\textbf{DY}: Beyond public respect, there is the issue of the proper
understanding of statistics as a competitive
discipline in the new age. Is it clear, for example, what core
knowledge should be required of
our graduate students? Are there standards for this that would have
wide appeal? If not, are there
consistent answers to the question of what we are all about? There is a
decent sense of where
we''ve come from, maybe much less of where statistics heads.

Which again brings up our history. It would seem that a lot more could
be preserved of the story of the growth of
Statistics over the past 100 years, and a sense of the people who
propelled this. On the positive side one sees a growing interest in
doing something about it, the Agresti and Meng project is just one
example of this.

\textbf{JS}: Thinking about the future of the field should be done periodically,
even if lamely. I am struck by the sudden emergence of books and articles
(e.g., \textit{The Information} by James Glieck; the special
edition of \textit{Science}
dated 11 February 2011) about the data flood threatening to drown us
or drive
everybody nuts. Apart from the need to physically manage the data, the issue
of how to analyze them has enormous implications for developments in
the field,
many of which, of course, have been in progress for some time and in
critical ways
(one example: false discovery rates to manage multiplicity in bioinformatics).
Still, there is so much going on now that, say, doubling the number of
practicing statisticians would still leave unfilled needs.

\textbf{DY}: What concerns me then is not so much the progress of statistical
technology, call it the
Benthamite school as described in your Hazelwood paper with Paul Meier
and Sandy Zabell,
but the well-being of the ``strict constructivist''
agenda that claims the\vadjust{\goodbreak} other end of the spectrum. Proceeding on the
basis of ``what is useful is good''
allows much latitude for producing new procedures in the light of
mushrooming data sets
and increased computational pow\-er, but at the same time evaluation and
validation remain as understaffed pursuits.

Methodological advances clearly outpace their justification.
Accentuating this problem, from my perspective,
is the flood of research papers that look and sound the same: ``Here is
our new procedure and these
are its asymptotic properties; we have run some simulations and
analyzed some `real' data, all of which goes
to show that our procedure is better than the other procedures of this type.''
All too often, the data set employed is dated and well worked over, and
the immediate contribution
to overall understanding of the main issues is not demonstrably nonignorable.
Against this, one finds that model validation is important but hugely
difficult, and
evaluation of large and continuing issues of public welfare that rely
on statistical
information is nowhere near what it should be. Do many graduate
programs give
serious attention to validation and evaluation? These are tough
problems, but when the going gets tough\,\ldots.

You likely think in terms of a broader agenda for statistics that gets
toward public policy.
Does this fit in such a descriptive framework, as an extra leg perhaps?

\textbf{JS}: Yes, most definitely yes, an expanded engagement in evaluation and
validation should be part of the field's
agenda. Though model validation has surfaced as a critical area in
several communities, with programs of
``Uncertainty Quantification'' that bring out the usual suspects as well
as whole varieties of engineers and
scientist, uncertainty is as uncertain as ever. Related are issues of
evaluation: ``just what does this
series of studies/analyses imply?'' Engagement with these questions
(whether in health, environment, education, etc.)
is not for the faint-hearted and needs many replacements for David
Freedman with the ability and energy to tackle
such problems.

Beyond these needs, your comments raise, I think, an issue about how
statistical evidence and ``proof'' are evolving.
In the past, mathematical proof and assessment was primary. Today,
computer simulations, in some contexts where
mathematical argument is unavailable, offer a less austere route;
perhaps ``preponderance of evidence'' versus ``beyond
a~reasonable doubt.'' The tendency you note of producing a~method and
assessing its utility by applying it to a~shopworn data set ought to lead to the case being tossed out of
statistical court on grounds of insufficient evidence.
Presumably, the weight of evidence is increased if the application is
made to a spectrum of data sets buttressed by
simulation studies. How to devise the spectrum and studies for a ``prima
facie'' case is not apparent but surely
worth thinking about.

\textbf{DY}: A lot of issues on the plate, but we seem better equipped to look
back at this point.
Maybe we could reminisce a bit, we've covered a lot of ground in 50
plus years.
What do you think of in terms of the good and, perhaps, the bad for you?

\textbf{JS}: As with everybody there were triumphs and disappointments, wins and losses.
Still the feeling that lasts and continues to drive me is that I had,
and still have,
a part in an exciting trip over meaningful terrain, accompanied by good people
(and a couple of scoundrels). I sometimes feel sorry for colleagues in other
disciplines who don't have the opportunity to swing in whatever style
comes up,
whether it be education, materials science, genomics, lottery draws, climate,
baseball---you get what\break I mean.

\textbf{DY}: I found that working on statistics problems of every sort was
natural and
pleasurable for me. In general, though, the profession itself has
served as a comfort zone,
and the good of this starts with the people---mentors, students (and
especially one's Ph.D. students),
colleagues, collaborators, friends. The list is so long that it is best
left unrecorded.
OK, you. But seriously, why would one choose to be something other than
a statistician?

\end{document}